\documentclass[runningheads]{llncs}
\usepackage[T1]{fontenc}
\usepackage{graphicx}
\usepackage{amsmath}
\usepackage{amsfonts}
\usepackage{amssymb}
\usepackage{mathrsfs}
\usepackage{stmaryrd}
\usepackage{xcolor}
\usepackage{tikz}

\newcommand{\atm}{\mathit{Atm}}

\newcommand{\dec}{\mathit{Dec}}
\newcommand{\val}{\mathit{Val}}

\newcommand{\mcm}{\text{$\Gamma \hspace{-0.05cm} = \hspace{-0.05cm} (S, \Phi)$}}
\newcommand{\pmcm}{\text{$(\Gamma, s, f)$}}
\newcommand{\mdm}{\text{$M \hspace{-0.05cm} = \hspace{-0.05cm} (W, \sim_{\allins}, \sim_{\univ}, V)$}}

\newcommand{\tagLabel}[2]{\tag{\textbf{#1}}\label{#2}}

\renewcommand{\phi}{\varphi}

\newcommand{\allins}{\Box_{\mathtt{I}}}
\newcommand{\someins}{\Diamond_{\mathtt{I}}}
\newcommand{\univ}{\Box_{\mathtt{F}}}
\newcommand{\possuniv}{\Diamond_{\mathtt{F}}}

\newcommand{\bclu}{\text{$\textsf{PLC}$}}

\newcommand{\wbclu}{\textsf{WPLC}}

\newcommand{\dbclu}{\text{$\textsf{D-PLC}$}}
\newcommand{\dwbclu}{\text{$\textsf{D-WPLC}$}}

\newcommand{\pimp}{\text{$\mathtt{PImp}$}}

\newcommand{\takevalue}[1]{\mathsf{t}({#1})}

\newcommand{\axp}{\textsf{$\mathtt{AXp}$}}

\newcommand{\putaway}[1]{}

\newcommand{\conj}[2]{\mathsf{cn}_{#1{,}#2}}

\begin{document}
	\title{A Logic of ``Black Box'' Classifier Systems } 
	
	%
	\author{Xinghan Liu\inst{1} \and
		Emiliano Lorini\inst{2}}
	%
	
	%
	\institute{
		IRIT, University of Toulouse, France \and
		IRIT, CNRS, University of Toulouse, France 
	}
	\maketitle              
	\begin{abstract}
		Binary classifiers
		are traditionally studied by propositional logic ($\mathsf{PL}$). 
		$\mathsf{PL}$ can only represent them as white boxes,
		under the assumption that 
		the underlying Boolean function
		is fully known. 
		Binary classifiers used in practical applications
		and trained by machine learning
		are however opaque. They are usually described as black boxes.
		In this paper, we provide a product modal logic called \bclu \ 
		(Product modal Logic for binary input Classifier)
		in which the notion of 
		``black box'' is interpreted as the uncertainty over
		a set of classifiers.
		We give results
		about axiomatics and complexity of satisfiability checking
		for our logic. 
		Moreover, we present a dynamic extension
		in which the process of acquiring new information
		about the actual 
		classifier can be represented.  
		
	\end{abstract}
	\section{Introduction}
	The notions of explanation and explainability have been extensively investigated by philosophers
	\cite{hempel1948studies,kment2006counterfactuals}
	and are key aspects of AI-based systems given the importance of explaining the behavior and prediction of an artificial intelligent system. 
	A variety of notions of explanation 
	for classifier systems 
	have been discussed in the area of explainable AI (XAI).
	Since systems trained by machine learning are increasingly opaque,
	instead of studying specific models,
	the \emph{model-agnostic} approach comes into focus.
	Namely, given a black box system or algorithm, we know nothing about how it works inside.
	Without opening the black box,
	we can query some (but not all) inputs and have some partial information about the system.
	Initially there were \emph{global} model-agnostic explanations like partial dependence plots and global surrogate models.
	Recently LIME \cite{ribeiro2016should} and its followers e.g. SHAP \cite{lundberg2017unified} and Anchors \cite{ribeiro2018anchors} have raised a \emph{local} model-agnostic explanation approach, namely
	explaining why a given input is classified in a certain way. 
	For a comprehensive overview
	of the research in this area see, e.g., \cite{molnar2020interpretable}. 
	
	At the mathematical level, a binary classifier can be viewed as a Boolean function and is traditionally studied by propositional logic.
	Recent years have witnessed several logic-based approaches
	to local explanation
	of classifier systems \cite{shih2018formal,DBLP:conf/ecai/DarwicheH20,ignatiev2019abduction,ignatiev2020contrastive,audemard2021computational}, e.g., computing prime implicants
	and abductive explanations
	of a given classification, and detecting biases 
	in the classification process
	by means of the notion  counterfactual explanation.
	But,  all these logic-based
	approaches deal with ``white box'' classifiers, i.e., specific transparent models 
	representable
	by  propositional formulas.
	A limitation is that
	given a Boolean function $f$ and a propositional formula $\phi$, $\phi$ either fully expresses $f$ or does not express $f$ at all.
	This all-or-nothing nature makes it impossible to give a \emph{partial} description of $f$, which is a natural way to represent a black box
	classifier. 
	
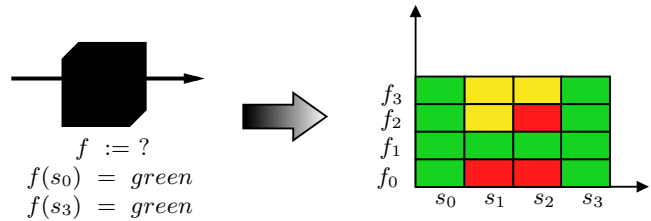
\begin{figure}
		\centering

		\begin{minipage}{1\linewidth}
			\centering
			
			
			\tikzset {_2qsx60nq2/.code = {\pgfsetadditionalshadetransform{ \pgftransformshift{\pgfpoint{0 bp } { 0 bp }  }  \pgftransformrotate{0 }  \pgftransformscale{2 }  }}}
			\pgfdeclarehorizontalshading{_cbpztmok4}{150bp}{rgb(0bp)=(0,0,0);
				rgb(37.767857142857146bp)=(0,0,0);
				rgb(62.5bp)=(1,1,1);
				rgb(100bp)=(1,1,1)}
			\tikzset{_aivza4or8/.code = {\pgfsetadditionalshadetransform{\pgftransformshift{\pgfpoint{0 bp } { 0 bp }  }  \pgftransformrotate{0 }  \pgftransformscale{2 } }}}
			\pgfdeclarehorizontalshading{_mcl3gggb1} {150bp} {color(0bp)=(transparent!10);
				color(37.767857142857146bp)=(transparent!10);
				color(62.5bp)=(transparent!0);
				color(100bp)=(transparent!0) } 
			\pgfdeclarefading{_x3ecq7gnr}{\tikz \fill[shading=_mcl3gggb1,_aivza4or8] (0,0) rectangle (50bp,50bp); } 
			\tikzset{every picture/.style={line width=0.5pt}} 
			
			\tikzset{every picture/.style={line width=0.70pt}} 
			
			\begin{tikzpicture}[x=0.45pt,y=0.45pt,yscale=-1,xscale=1]
				
				\draw    (373.83,168.26) -- (566,168) ;
				\draw [shift={(569,168)}, rotate = 179.92] [fill={rgb, 255:red, 0; green, 0; blue, 0 }  ][line width=0.08]  [draw opacity=0] (8.93,-4.29) -- (0,0) -- (8.93,4.29) -- cycle    ;
				\draw    (373.83,168.26) -- (373.83,19) ;
				\draw [shift={(373.83,16)}, rotate = 90] [fill={rgb, 255:red, 0; green, 0; blue, 0 }  ][line width=0.08]  [draw opacity=0] (8.93,-4.29) -- (0,0) -- (8.93,4.29) -- cycle    ;
				\draw  [fill={rgb, 255:red, 248; green, 231; blue, 28 }  ,fill opacity=1 ] (497.05,75.32) -- (497.05,98.56) -- (455.98,98.56) -- (455.98,75.32) -- cycle ;
				\draw  [fill={rgb, 255:red, 248; green, 231; blue, 28 }  ,fill opacity=1 ] (455.98,75.32) -- (455.98,98.56) -- (414.9,98.56) -- (414.9,75.32) -- cycle ;
				\draw  [fill={rgb, 255:red, 23; green, 211; blue, 33 }  ,fill opacity=1 ] (414.9,75.32) -- (414.9,98.56) -- (373.83,98.56) -- (373.83,75.32) -- cycle ;
				\draw  [fill={rgb, 255:red, 23; green, 211; blue, 33 }  ,fill opacity=1 ] (414.9,98.56) -- (414.9,121.79) -- (373.83,121.79) -- (373.83,98.56) -- cycle ;
				\draw  [fill={rgb, 255:red, 248; green, 231; blue, 28 }  ,fill opacity=1 ] (455.98,98.56) -- (455.98,121.79) -- (414.9,121.79) -- (414.9,98.56) -- cycle ;
				\draw  [fill={rgb, 255:red, 255; green, 27; blue, 27 }  ,fill opacity=1 ] (497.05,98.56) -- (497.05,121.79) -- (455.98,121.79) -- (455.98,98.56) -- cycle ;
				\draw  [fill={rgb, 255:red, 23; green, 211; blue, 33 }  ,fill opacity=1 ] (414.9,121.79) -- (414.9,145.03) -- (373.83,145.03) -- (373.83,121.79) -- cycle ;
				\draw  [fill={rgb, 255:red, 23; green, 211; blue, 33 }  ,fill opacity=1 ] (455.98,121.79) -- (455.98,145.03) -- (414.9,145.03) -- (414.9,121.79) -- cycle ;
				\draw  [fill={rgb, 255:red, 23; green, 211; blue, 33 }  ,fill opacity=1 ] (497.05,121.79) -- (497.05,145.03) -- (455.98,145.03) -- (455.98,121.79) -- cycle ;
				\draw  [fill={rgb, 255:red, 23; green, 211; blue, 33 }  ,fill opacity=1 ] (414.9,145.03) -- (414.9,168.26) -- (373.83,168.26) -- (373.83,145.03) -- cycle ;
				\draw  [fill={rgb, 255:red, 255; green, 27; blue, 27 }  ,fill opacity=1 ] (455.98,145.03) -- (455.98,168.26) -- (414.9,168.26) -- (414.9,145.03) -- cycle ;
				\draw  [fill={rgb, 255:red, 255; green, 27; blue, 27 }  ,fill opacity=1 ] (497.05,145.03) -- (497.05,168.26) -- (455.98,168.26) -- (455.98,145.03) -- cycle ;
				\draw  [fill={rgb, 255:red, 23; green, 211; blue, 33 }  ,fill opacity=1 ] (537.24,75.32) -- (537.24,98.56) -- (496.16,98.56) -- (496.16,75.32) -- cycle ;
				\draw  [fill={rgb, 255:red, 23; green, 211; blue, 33 }  ,fill opacity=1 ] (537.24,98.56) -- (537.24,121.79) -- (496.16,121.79) -- (496.16,98.56) -- cycle ;
				\draw  [fill={rgb, 255:red, 23; green, 211; blue, 33 }  ,fill opacity=1 ] (537.24,121.79) -- (537.24,145.03) -- (496.16,145.03) -- (496.16,121.79) -- cycle ;
				\draw  [fill={rgb, 255:red, 23; green, 211; blue, 33 }  ,fill opacity=1 ] (537.24,145.03) -- (537.24,168.26) -- (496.16,168.26) -- (496.16,145.03) -- cycle ;
				\draw  [fill={rgb, 255:red, 0; green, 0; blue, 0 }  ,fill opacity=1 ] (77,60.67) -- (90.67,47) -- (147,47) -- (147,103.33) -- (133.33,117) -- (77,117) -- cycle ; \draw   (147,47) -- (133.33,60.67) -- (77,60.67) ; \draw   (133.33,60.67) -- (133.33,117) ;
				\draw [line width=1.5]    (105,77) -- (191,77) ;
				\draw [shift={(195,77)}, rotate = 180] [fill={rgb, 255:red, 0; green, 0; blue, 0 }  ][line width=0.08]  [draw opacity=0] (15.6,-3.9) -- (0,0) -- (15.6,3.9) -- cycle    ;
				\draw [line width=1.5]    (34,77) -- (102,77) ;
				\draw [shift={(105,77)}, rotate = 180] [color={rgb, 255:red, 0; green, 0; blue, 0 }  ][line width=1.5]    (14.21,-4.28) .. controls (9.04,-1.82) and (4.3,-0.39) .. (0,0) .. controls (4.3,0.39) and (9.04,1.82) .. (14.21,4.28)   ;
				\path  [shading=_cbpztmok4,_2qsx60nq2,path fading= _x3ecq7gnr ,fading transform={xshift=2}] (229,99) -- (271,99) -- (271,89) -- (299,109) -- (271,129) -- (271,119) -- (229,119) -- cycle ; 
				\draw   (229,99) -- (271,99) -- (271,89) -- (299,109) -- (271,129) -- (271,119) -- (229,119) -- cycle ; 

				\draw (338.5,149.77) node [anchor=north west][inner sep=0.75pt]    {$f_{0}{}$};
				\draw (341.11,124.07) node [anchor=north west][inner sep=0.75pt]    {$f_{1}$};
				\draw (341.55,100.34) node [anchor=north west][inner sep=0.75pt]    {$f_{2}$};
				\draw (373.95,171.52) node [anchor=north west][inner sep=0.75pt]    {$\ \ s_{0}{}$};
				\draw (415.4,171.52) node [anchor=north west][inner sep=0.75pt]    {$\ \ s_{1}$};
				\draw (456.48,171.52) node [anchor=north west][inner sep=0.75pt]    {$\ \ s_{2}$};
				\draw (342,79.58) node [anchor=north west][inner sep=0.75pt]    {$f_{3}$};
				\draw (497.55,171.52) node [anchor=north west][inner sep=0.75pt]    {$\ \ s_{3}$};
				\draw (41,121.4) node [anchor=north west][inner sep=0.75pt]    {$ \begin{array}{l}
						\ \ \ \ \ \ f\ :=\ ?\\
						f( s_{0}) \ =\ green\\
						f( s_{3}) \ =\ green
					\end{array}$};

			\end{tikzpicture}
		\end{minipage}
		
		\caption{A classifier associating color labels
			in  $\{\textrm{red, yellow, green}\}$
			to input instances.
			We do not know its  Boolean formula,
			since $f_0, f_1, f_2, f_3$ are all 
			compatible with our partial knowledge of it. 
			However, we know that the two input instances $s_0$ and $s_3$ are both 
			classified as green.
		}
		\label{fig: PLC black box}
	\end{figure}

	The central idea of this paper is that
	a product modal logic is
	the 
	proper
	way 
	to represent a ``black box'' classifier.
	As we have shown in \cite{LiuLorini2021BCL},
	it is natural to think of a classifier
	with binary inputs 
	as a partition of an S5 Kripke model, where each possible state stands for an input instance.
	However, this only represents ``white box'' classifiers.
	We extend this
	semantics with a second dimension  universally ranging over 
	a set of  possible classifiers, which results in 
	a proper extension
	of the product modal logic $S5\times S5= S5^2 $
	\cite{GabbayWolther}
	we call $\bclu$ (Product modal Logic
	of binary-input Classifiers). 
	The notion of
	black box is interpreted as an  agent's uncertainty among those (white box) classifiers, as illustrated in 
	Figure  \ref{fig: PLC black box}.
	
	The paper is structured as follows.
	Section 2 introduces the modal language and semantic model of $\bclu$ which we name multi-classifier model (MCM). 
	Its axiomatics along with the completeness and complexity results
	for the satisfiability checking problem
	are given in Section 3.
	In Section 4,
	we will exemplify the logic's application by  using it to represent 
	the notion of black box and  to
	formalize 
	different notions of classifier explanation. 
	A dynamic extension is given in Section 6 to capture the
	process of acquisition
	of new knowledge about
	the classifier. 
	Some non-routine proofs are given in 
	a technical annex at the end
	of the paper.

	\section{Language and Semantics}
	Let $\atm_0=\{p, q, \ldots \}$ be a countable set of atomic propositions which intend to denote input variables (features) of a classifier.
	We introduce a finite set $\val$ to denote the possible output values (classifications, decisions) of the classifier.
	Elements of $\val$ are noted $x, y, \ldots $
	For any $x \in \val$, we call $\takevalue{x}$ a decision atom, and have $\dec = \{\takevalue{x} : x \in \val\}$.\footnote{Notice that $p$ denotes an input \emph{variable}, while $x$ is an output \emph{value} rather than the output \emph{variable}, which makes sense of the symbolic difference between $p$ and $\takevalue{x}$.}
	Finally let $\atm = \atm_0 \cup \dec$.
	

	The modal language  $\mathcal{L} $
	is defined by the following grammar:
	\begin{center}\begin{tabular}{lcl}
			$\varphi$  & $::=$ & $ p \mid
			\takevalue{x} \mid 
			\neg\varphi \mid \varphi_1\wedge\varphi_2 \mid \allins\varphi \mid {\univ} \phi,$
	\end{tabular}\end{center}
	where $p$ ranges over $ \atm_0 $
	and
	$x $
	ranges over $\val $.
	
	\begin{definition}\label{def:MCM}
		A multi-classifier model (MCM) is a pair $\mcm$ where $S \subseteq 2^{\atm_0}$ and
		$\Phi \subseteq  F_S$,  with
		$F_S =\val^S$ 
		the set of functions
		with domain $S$
		and codomain $\val$. 
		A pointed MCM is a triple 
		$( \Gamma ,  s,f  ) $
		where 
		$\mcm$
		is
		an MCM,
		$s \in S$
		and $f \in \Phi$. 
		The class of all multi-classifier models is noted $ \mathbf{MCM}$. 
	\end{definition}
	
	
	Formulas in 
	$\mathcal{L} $ are interpreted
	relative
	to a pointed MCM as follows. 
	\begin{definition}[Satisfaction relation]\label{truthcond1}
		Let 
		$\mcm$ be an MCM,
		$s \in S$
		and $f \in \Phi$. Then, 
		\begin{eqnarray*}
			( \Gamma ,  s,f  )
			\models p & \iff & p \in s, \\
			( \Gamma ,  s,f  ) \models \takevalue{x} & \iff & f(s) = x,  \\
			( \Gamma ,  s,f  ) \models \neg \phi & \iff & ( \Gamma ,  s,f  ) \not \models \phi, \\
			( \Gamma ,  s,f  )
			\models \phi \wedge \psi & \iff & ( \Gamma ,  s,f  ) \models \phi \text{ and } ( \Gamma ,  s,f  ) \models \psi, \\
			( \Gamma ,  s,f  ) \models \allins \phi & \iff & \forall s' \in S: ( \Gamma ,  s',f  ) \models \phi, \\
			( \Gamma ,  s,f  )
			\models {\univ} \phi & \iff & \forall f' \in \Phi : ( \Gamma ,  s,f' )
			\models \phi .
		\end{eqnarray*}
	\end{definition}
	Both $\allins \phi$ and $\univ \phi$ have standard modal reading but range over different sets.
	$\allins \phi$
	has to be read ``$\phi$ \emph{necessarily} holds
	for the actual function, regardless
	of the input instance'', while its dual $\someins \phi =_{\mathit{def}} \neg \allins \neg \phi$ has to be read  ``$\phi$ \emph{possibly} holds
	for the actual function, regardless
	of the input instance''.
	Similarly, $\univ \phi$ has to be read ``$\phi$ \emph{necessarily} holds
	for the actual input instance, regardless
	of the function'' and its dual $\possuniv \phi$ has to be read ``$\phi$ \emph{possibly} holds
	for the actual input instance, regardless
	of the function''.

	

	Let $X$
	be a finite subset of 
	$\atm_0$.
	An important abbreviation is the following:
	\begin{align*}
		[X] \varphi =_{\mathit{def}} &   \bigwedge_{Y \subseteq X} \big( (\bigwedge_{p \in Y} \wedge \bigwedge_{p \in X \setminus Y} \neg p) \to
		\allins ( (\bigwedge_{p \in Y} \wedge \bigwedge_{p \in X \setminus Y} \neg p) \to \phi)\big).
	\end{align*}
	Complex as it seems, $[X]\phi$ means nothing but ``$\phi$ necessarily holds, regardless of the values of the input variables outside $X$''
	or 
	``$\phi$ necessarily holds, 
	if the values of the input variables in $X$ are kept fixed''.  
	It can be justified by checking that $\pmcm \models [X] \phi$, if and only if $\forall s' \in S,$ if $s \cap X = s' \cap X$ then $(\Gamma, s', f) \models \phi$.
	Its dual $\langle X \rangle \varphi  =_{\mathit{def}} \neg [X] \neg \phi$ has to be read ``$\phi$ \emph{possibly} holds, if the values of the input variables in $X$ are kept fixed'''. 
	These modalities  have a \emph{ceteris paribus} reading and were first introduced in \cite{LoriniCETERISPARIBUS}. Similar modalities are used in existing  logics of functional dependence   \cite{YangVaananen2016,DBLP:journals/jphil/BaltagB21}.

	
	A formula $\varphi$
	of
	$\mathcal{L} $
	is said to 
	be satisfiable
	relative
	to the class 
	$\mathbf{MCM}$
	if there exists
	a pointed multi-classifier model 
	$(\Gamma,s, f)$
	with $\Gamma \in \mathbf{MCM} $
	such that $(\Gamma, s, f) \models \varphi$.
	We say that that $\varphi$
	is valid in the multi-classifier 
	model
	$\mcm$, noted
	$\Gamma \models \phi$,
	if $(\Gamma, s, f) \models \varphi$
	for every $s \in S, f \in \Phi$.
	It is said to be valid
	relative to 
	$\mathbf{MCM}$,
	noted $\models_{\mathbf{MCM}} \varphi$,
	if
	$\neg \varphi$
	is not satisfiable relative
	to $\mathbf{MCM}$.

	\section{Axiomatics and Complexity}

	In this section,
	we are going  to present
	two
	axiomatics
	for the language
	$\mathcal{L} $
	by distinguishing the finite-variable
	from
	the infinite-variable case.
	We
	will moreover give complexity results
	for satisfiability checking. 
	Before, 
	we are going to introduce 
	an alternative Kripke semantics
	for 
	the interpretation of the language 
	$\mathcal{L} $.
	It will allow us to 
	use the standard canonical model technique for proving completeness. Indeed, this technique
	cannot be directly applied to MCMs in the infinite-variable case.

	\subsection{Alternative Kripke Semantics}
	The crucial concept
	of the  alternative semantics
	is multi-decision model (MDM).

	\begin{definition}\label{MDM}
		An MDM is a tuple 
		$M= \big(W, \sim_{\allins},
		\sim_{\univ} , V\big)$
		where:
		\begin{itemize}
			\item[-] 
			$W$ is a set of worlds,
			\item[-]
			$\sim_{\allins}$ and $ \sim_{\univ}$
			are equivalence
			relations on $W$, 
			\item[-] 
			$V: W \longrightarrow
			2^{\mathit{Atm}}$
			is a valuation function, 
		\end{itemize}
		and which satisfies the following constraints,
		$\forall w,v \in W$,
		$\forall x, y \in \val :$
		\begin{description}
			\item[($\mathbf{C1}$)] 
			$ \sim_{\allins} \circ \sim_{\univ}=
			\sim_{\univ} \circ \sim_{\allins} $, 
			
			
			\item[($\mathbf{C2}$)] 
			if
			$V_{\atm_0}(w) = V_{\atm_0}(v)$
			and $w \sim_{\allins} v$,
			then $V_\dec(w) = V_\dec(v)$,
			
			\item[($\mathbf{C3}$)] 
			if $w  \sim_{\univ} v$
			then $V_{\atm_0}(w)=V_{\atm_0}(v)$,
			
			\item[($\mathbf{C4}$)] 
			if $\takevalue{x} \in V(w)$
			and $x \neq y$
			then 
			$\takevalue{y} \not  \in V(w)$, 
			
			\item[($\mathbf{C5}$)] 
			$\exists x \in
			\val \text{ such that }\takevalue{x} \in V(w)$,
		\end{description}
		with  $V_Y(w)= \big( V(w) \cap Y\big)$
		for every
		$w \in W$ and for every $Y \subseteq  \atm $, 
		and $\circ$
		the standard composition operator
		for binary relations. 
		
	\end{definition}
	The class of multi-decision models
	is noted $\mathbf{MDM}$.
	An MDM \mdm\
	is called finite if $W$
	is finite. The class of finite MDMs is noted 
	finite-$\mathbf{MDM}$.
	Interpretation of formulas
	in 
	$ \mathcal{L} $
	relative
	to a pointed MDM 
	goes as follows. (We omit interpretations for
	$\neg$ and $\wedge $ which are defined as usual.)
	\begin{definition}[Satisfaction Relation]\label{truthcond2}
		Let     $M= \big(W, \sim_{\allins},
		\sim_{\univ} , V\big)$
		be an MDM and let $w \in W$.
		Then, 
		\begin{eqnarray*}
			(M, w) \models q & \Longleftrightarrow & 
			q \in 
			V(w) \text{ for }q \in\mathit{Atm} , \\
			(M,w) \models \allins \varphi
			& \Longleftrightarrow & 
			\forall v \in W,
			\text{ if }
			w \hspace{-0.05cm} \sim_{\allins} \hspace{-0.05cm} v 
			\text{ then } v \models \varphi,\\
			(M, w) \models {\univ} \phi & \Longleftrightarrow & \forall v \in W, \text{if } w \sim_{\univ}  v \text{ then } v \models \phi.
		\end{eqnarray*}
	\end{definition}

	Validity and satisfiability
	of formulas
	in $\mathcal{L} $
	relative to   class  $\mathbf{MDM}$ (resp.   finite-$\mathbf{MDM}$) are  defined in the usual way.

	The most important result in this subsection is the semantic  equivalence between $\mathbf{MCM}$ and $\mathbf{MDM}$, regardless of $\atm_0$ being finite or infinite.
	Although a pointed MDM $(M, w)$ looks
	like a pointed MCM $(\Gamma, s, f)$,
	it only approximates it. 
	Indeed, unlike 
	an 
	MCM,
	an 
	MDM $M$ may be redundant, that is, 
	(i) a classifier in $M $
	(i.e., a $\sim_{\allins}$-equivalence class)
	may include multiple copies
	of the same input instance 
	(i.e., 
	of the same valuation
	for the atoms in $\atm_0$), or 
	(ii) $M $
	may contain 
	multiple copies of the same classifier
	(i.e., two identical 
	$\sim_{\univ}$-equivalence classes).
	Moreover, an MDM
	$M$ may be ``defective''
	insofar as (iii) the intersection
	between a classifier in $M$
	(i.e., a $\sim_{\allins}$-equivalence class)
	and
	the set of all possible
	classifications
	of a given input instance
	by the classifiers in $M$
	(i.e., a $\sim_{\univ}$-equivalence class)
	is not a singleton.   
	What  makes the proof
	of the following theorem  non-trivial
	is transforming a possibly
	redundant or defective MDM
	into a non-redundant and non-defective one
	by preserving truth of formulas. 
	A non-redundant and non-defective MDM
	is then isomorphic to an MCM. 
	
	\begin{theorem}\label{theo:equiv1}
		Let $\varphi \in \mathcal{L}  $.
		Then, $\varphi$
		is satisfiable relative to the
		class 
		$\mathbf{MCM}$
		if and only if it is satisfiable
		relative
		to the class
		$\mathbf{MDM}$.
	\end{theorem}
	\begin{proof}
		We start with the left-to-right direction
		of the proof.
		Let 
		$( \Gamma ,  s_0,f_0  ) $ be a pointed MCM
		with 
		$\mcm$, 
		$S \subseteq 2^{\atm_0}$ and
		$\Phi \subseteq  F_S$
		such that 
		$( \Gamma ,  s_0,f_0  ) \models \varphi$.
		We
		define
		the tuple 
		$M= \big(W, \sim_{\allins},
		\sim_{\univ}, V\big)$
		as follows:
		\begin{itemize}
			\item[-] $W= \{(s,f) : s \in S \text{ and }
			f \in \Phi
			\} $, 
			\item[-]
			$\forall (s, f), (s', f') \in W$, $(s, f) \sim_{\allins} (s', f')$ iff $f = f'$
			
			\item[-]
			$ \forall (s,f), (s',f') \in W$,
			$(s,f)  \sim_{\univ} (s',f' )$
			iff $ s=s'$,
			
			\item[-]
			$\forall  (s,f)  \in W$,
			$V(s,f) = s \cup \{ 
			\takevalue{f(s)} 
			\} $.

		\end{itemize}
		It is routine
		exercise to verify 
		that 
		$M$
		so defined is an MDM.
		Moreover, by induction on the structure
		of $\varphi$, 
		it is easy to prove  that ``$( \Gamma ,  s,f  ) \models \varphi$ iff 
		$\big( M,  (s,f)  \big) \models \varphi$''
		for every 
		$s \in S$
		and 
		$f \in \Phi$. 
		Thus, 
		$\big( M,  (s_0,f_0)  \big) \models \varphi$
		since $( \Gamma ,  s_0,f_0  ) \models \varphi$.
		
		Let us now prove
		the right-to-left direction. 
		Let $\mdm$
		be an MDM
		and
		$w_0 \in W$
		such that $(M,w_0)\models \varphi $.
		Given $v \in W $,
		let 
		$|v|= \{ u \in W: v
		\sim_{\allins}
		u \text{ and }
		V(v)=V (u) \}$. 
		We transform 
		the MDM $M$
		into a tuple 
		$M'= \big(W', \sim_{\allins}',
		\sim_{\univ}' , V' \big)$
		such that:
		\begin{itemize}
			\item[-] $W'= \{  
			|v|: v \in W
			\}$, 
			\item[-] 
			$ \forall 
			|v|, 
			|u| \in W'$, 
			$|v|
			\sim_{\allins}
			|u| $ 
			iff
			$\exists v' \in  |v|,
			u' \in |u|$
			such that $v' \sim_{\allins} u' $,

			\item[-] 
			$ \forall 
			|v|, 
			|u| \in W'$, 
			$|v|
			\sim_{\univ}'
			|u| $ 
			iff
			$\exists v' \in  |v|,
			u' \in |u|$
			such that $v'   \sim_{\univ} u' $, 
			
			\item[-]
			$ \forall 
			|v| \in W'$, 
			$V' (  |v|)=
			V (v) $. 
			
		\end{itemize}
		Like what we did for $V$, let $V'_Y(|v|) = V'(|v|) \cap Y$ for all $Y \subseteq \atm$.

		It is a routine exercise to verify that 
		$M'$
		is an MDM and, by induction
		on the structure of $\varphi$,
		that 
		``$( M,v  ) \models \varphi$ iff 
		$(M',|v| )\models \varphi$''
		for every 
		$v \in W$.
		Thus, 
		$(M,|w_0| )\models \varphi $
		since  $(M,w_0)\models \varphi $.
		Finally, because
		of Constraints $\mathbf{C2}$ and
		$\mathbf{C3}$
		in Definition
		\ref{MDM},
		the following  property 
		holds:
		\begin{align*}
			\mathbf{(C6)} \ \ \ \big( \sim_{\allins}' \cap \sim_{\univ}'
			\big) =
			\mathit{id}_{W'}, 
		\end{align*}
		where 
		$ \mathit{id}_{W'}$
		is the identity relation on $W'$.

		Let $W' /  \sim_{\allins}' $
		be the quotient set of $W'$
		by the equivalence relation $\sim_{\allins}'$.
		We note $\tau, \tau', \ldots $ its elements. 
		Given $\tau, \tau' \in W' /  \sim_{\allins}'$,
		we write
		$\tau \approx_{F} \tau' $
		if and only if,
		$ \forall |v| \in  \tau, \forall  |u| \in \tau'$,
		if $V_{\atm_0}'(|v|)=
		V_{\atm_0}'(|u|)$
		then 
		$V_{\dec}'(|v|)=
		V_{\dec}'(|u|)$.
		Given $|v|, |u| \in W'$,
		we write
		$|v| \simeq |u| $
		if and only if $\exists \tau, \tau' \in 
		W' /  \sim_{\allins}'$
		such that 
		$|v|\in \tau ,
		|u| \in \tau'$, $\tau \approx_{F} \tau' $
		and $V_{\atm_0}' (|v|)  =V_{\atm_0}'(|u|)$. 
		Clearly, 
		$\approx_{F}$
		and
		$\simeq$
		are equivalence relations. 
		
		We are going to transform 
		the MDM $M'$
		into an
		MDM
		which does not contain multiple copies
		of the same
		function
		and which satisfies the same
		formulas
		as $M'$.
		We define it to be 
		a tuple 
		$M''= \big(W'', \sim_{\allins}'',
		\sim_{\univ}'' , V'' \big)$
		such that:
		\begin{itemize}
			\item[-] $W''= \{  
			\simeq\!\!( |v|)  : |v| \in W'
			\}$, 
			
			\item[-] 
			$ \forall 
			\simeq\!\!( |v|), 
			\simeq\!\!( |u|) \in W''$, 
			$ \simeq\!\!( |v|)
			\sim_{\allins}''
			\simeq\!\!( |u|) $ 
			iff
			$\exists |v' | \in   \simeq\!\!( |v|),
			|u'| \in  \simeq\!\!( |u|  )$
			such that $|v'|  \sim_{\allins}' |u'| $, 
			
			\item[-] 
			$ \forall 
			\simeq\!\!( |v|), 
			\simeq\!\!( |u|) \in W''$, 
			$ \simeq\!\!( |v|)
			\sim_{\univ} ''
			\simeq\!\!( |u|) $ 
			iff
			$\exists |v'| \in   \simeq\!\!( |v|),
			|u'| \in  \simeq\!\!( |u|)$
			such that $|v'|  \sim _{\univ}' |u'|$, 
			
			\item[-]
			$ \forall 
			\simeq\!\!( |v|) \in W''$, 
			$V'' \big (   
			\simeq\!\!( |v|)
			\big )=
			V' (|v|) $. 
			
		\end{itemize}
		
		Again, it is routine  to verify that 
		$M''$
		is an MDM
		which  satisfies the previous
		Constraint $\mathbf{C6}$.
		Moreover, by induction
		on the structure of $\varphi$,
		it is easy to prove 
		that 
		``$(M',|v| ) \models \varphi$ iff 
		$\big(M'', \simeq\!\!( |v| ) \big)\models \varphi$''
		for every 
		$|v| \in W'$.
		Thus, 
		$\big(M'',
		\simeq\!\!( |w_0|) \big)\models \varphi $
		since 
		$(M',|w_0|)\models \varphi $.
		
		We can easily build an MCM isomorphic to $M''$.

	\end{proof}

	\subsection{Finite-Variable Case}
	
	We first
	consider
	the variant of the logic
	with finitely many propositional atoms
	in 
	$\atm_0$. 
	For every finite $X, Y  \subseteq \atm_0$
	we define:
	\begin{align*}
		\conj{X}{Y} =_{\mathit{def}} \bigwedge_{p \in X} p \wedge \bigwedge_{p \in (Y \setminus X)} \neg p.
	\end{align*}

	\begin{definition}[Logic $\bclu$]\label{axiomatics}
		Let $\atm_0$
		be finite.
		We define  $\bclu$
		as the extension of classical
		propositional logic given by 
		axioms and rules of inference in Table \ref{tab: bcl axioms}.
		
	\end{definition}
	
	\begin{table}[ht]
		\centering
		\small
		\begin{align}
			& \big( \blacksquare \varphi
			\wedge \blacksquare (\varphi \rightarrow \psi) \big)
			\rightarrow \blacksquare \psi
			\tagLabel{K$_{\blacksquare}$}{ax:Ksq}\\
			& \blacksquare  \varphi
			\rightarrow  \varphi
			\tagLabel{T$_{\blacksquare}$}{ax:Tsq}\\
			& \blacksquare \varphi
			\rightarrow  \blacksquare\blacksquare \varphi
			\tagLabel{4$_{\blacksquare}$}{ax:4sq}\\
			& \neg  \blacksquare \varphi
			\rightarrow  \blacksquare \neg  \blacksquare \varphi
			\tagLabel{5$_{\blacksquare}$}{ax:5sq}\\
			&  \univ \allins \varphi \leftrightarrow 
			\allins    \univ \phi
			\tagLabel{Comm}{ax:Comm}\\
			&  \bigvee_{x \in \val} \takevalue{x}
			\tagLabel{AtLeast$_{\takevalue{x} }$}{ax:AtLeast}\\
			&  \takevalue{x} \to \neg \takevalue{y} \text{ if }
			x \neq y
			\tagLabel{AtMost$_{\takevalue{x} }$}{ax:AtMost}\\
			&   \big(\conj{X}{\atm_0} \wedge \takevalue{x} \big) \to \allins \big(\conj{X}{\atm_0} \to \takevalue{x}\big)
			\tagLabel{Funct}{ax:Funct}\\
			&   p \rightarrow {\univ}  p
			\tagLabel{Indep$_{ {\univ},p }$}{ax:Ind1}\\
			&  \neg  p \rightarrow {\univ}  \neg  p
			\tagLabel{Indep$_{ {\univ},\neg p }$}{ax:Ind2}\\
			&  \frac{\varphi }{\blacksquare \varphi }
			\tagLabel{Nec$_{\blacksquare}$}{ax:NecK}
		\end{align}
		\caption{Axioms and rules of inference, with $\blacksquare \in \{\allins, {\univ}\}$}
		\label{tab: bcl axioms}
	\end{table}

	Axioms \ref{ax:AtLeast}, \ref{ax:AtMost} and \ref{ax:Funct}
	guarantee that every input $Y \subseteq \atm_0$, whose syntactic counterpart  is $\conj{Y}{\atm_0}$, has only one decision atom as output.
	Axioms
	\ref{ax:Ksq},
	\ref{ax:Tsq},
	\ref{ax:4sq}
	and
	\ref{ax:5sq}
	together with the rule
	of inference \ref{ax:NecK}
	indicate that
	both modal operators
	${\univ}$
	and $\allins$
	satisfy
	the principles of the modal logic S5.
	According to 
	Axiom
	\ref{ax:Comm}, they   moreover  commute.
	This makes the logic meet the requirement of a product of two S5 modal logics, i.e., $\textrm{S5}^2$  \cite{GabbayWolther}.
	Nevertheless, the existence of the
	two  ``independence'' Axioms \ref{ax:Ind1}
	and \ref{ax:Ind2}
	indicates that $\bclu$ is stronger than $\textrm{S5}^2$ in general.
	Soundness of \bclu\ relative to $\mathbf{MCM}$ is a simple exercise. 
	To prove the completeness result, we first need to show that  \bclu\ is complete relative to $\mathbf{MDM}$, which is proven by the canonical model construction.
	
	\begin{theorem}\label{theor: Completeness Finite}
		Let $\atm_0$ be finite.
		Then, the logic $\bclu$ is sound and complete relative
		to the class 
		$\mathbf{MDM}$.
	\end{theorem}
	
	Our main result of this subsection becomes a corollary of Theorems \ref{theo:equiv1} and \ref{theor: Completeness Finite}.
	
	\begin{corollary}\label{corollComp}
		Let $\atm_0$ be finite. Then, the logic \bclu\ is sound and complete relative to the class $\mathbf{MCM}$.
	\end{corollary}

	\subsection{Infinite-Variable Case}

	We now move to 
	the 
	infinite-variable variant of our logic,
	under the assumption 
	that the set  $\atm_0$
	is countably infinite. 
	In  order to obtain an axiomatics
	we just need to drop
	the 
	``functionality''
	Axiom \ref{ax:Funct}
	of Table \ref{tab: bcl axioms}.
	Indeed, 
	when 
	$\atm_0$ is infinite,
	the  construction 
	$\conj{X}{\atm_0}$  
	cannot be expressed  in a finitary way.
	
	\begin{definition}[Logic $\wbclu$]\label{axiomatics2}
		We define  
		$\wbclu$ (Weak $\bclu$)
		to be  the extension of classical
		propositional logic given by
		Axioms 
		\ref{ax:Ksq},
		\ref{ax:Tsq},
		\ref{ax:4sq},
		\ref{ax:5sq},
		\ref{ax:Comm},
		\ref{ax:AtLeast},
		\ref{ax:AtMost},
		\ref{ax:Ind1}
		and \ref{ax:Ind2},
		and the rule of inference \ref{ax:NecK} in Table \ref{tab: bcl axioms}.
	\end{definition}

	Soundness
	of the logic
	$\wbclu$
	is a straightforward exercise.
	For completeness, 
	we need to distinguish 
	MDMs from
	quasi-MDMs that are obtained by removing
	the ``functionality''
	Constraint $\mathbf{C2}$
	from Definition \ref{MDM}. 
	
	\begin{definition}[Quasi-MDM]\label{QMDM}
		A quasi-MDM is 
		a tuple
		$\mdm$
		where
		$W$,
		$\sim_{\allins}$,
		$\sim_{\univ}$
		and $V$
		are defined as in Definition
		\ref{MDM}
		and which satisfies all constraints
		of Definition 
		\ref{MDM} except $\mathbf{C2}$.
	\end{definition}
	The class of quasi-MDMs
	is noted $\mathbf{QMDM}$.
	A  quasi-MDM $\mdm$
	is said to be finite
	if $W$ is finite.
	The class of finite quasi-MDMs
	is noted finite-$\mathbf{QMDM}$.
	Semantic interpretation of formulas
	in 
	$ \mathcal{L} $
	relative to quasi-MDMs
	is analogous to semantic
	interpretation relative
	to MDMs given in Definition
	\ref{truthcond2}. Moreover, 
	validity and satisfiability
	of formulas
	in $\mathcal{L}  $
	relative to   class  $\mathbf{QMDM}$ (resp. finite-$\mathbf{QMDM}$) is again defined in the usual way.

	The first crucial result
	of this subsection
	is that when $\atm_0$
	is infinite the language 
	$ \mathcal{L}  $
	cannot distinguish finite MDMs from
	finite 
	quasi-MDMs. 
	
	\begin{theorem}\label{theo:equiv2}
		Let $\varphi \in \mathcal{L}  $
		with $\atm_0 $
		infinite. 
		Then, $\varphi$
		is satisfiable relative to the
		class 
		finite-$\mathbf{MDM}$
		if and only if it is satisfiable
		relative
		to the class  finite-$\mathbf{QMDM}$.
	\end{theorem}
	\begin{proof}
		The left-to-right direction is trivial.
		We prove the right-to-left direction.
		Suppose $\atm_0 $ is
		infinite.
		Moreover, let $\mdm$
		be a finite quasi-MDM
		and $w_0 \in W$ such that 
		$(M,w_0)\models \varphi $. 
		Since  $\atm_0 $
		is infinite and
		$W$ is finite, we can define
		an injection 
		$g : W \longrightarrow \atm_0 \setminus
		\atm(\varphi)$.
		We define the tuple 
		$M'= \big(W', \sim_{\allins}',
		\sim_{\univ}' , V'\big)$
		as follows:
		\begin{itemize}
			\item[-] $W'= W$;
			\item[-] $\sim_{\allins}' = \sim_{\allins}$
			\item[-] $\sim_{\univ}'= \sim_{\univ}$;
			
			\item[-] for every $w \in W'$,
			\begin{align*}
				V'(w) = \big( V(w) \setminus 
				\{g(v): v \in W \text{ and } w\neq v \}
				\big) \cup \{g(w)\}. 
			\end{align*}
		\end{itemize}
		It is routine  to verify that $M'$
		is a finite MDM. Indeed, $V_{\atm_0}'(w)\neq V_{\atm_0}'(v)$
		for all
		$w,v \in W'$
		such that $w \neq v$. This guarantees that $M'$
		satisfies the ``functionality'' constraint $\mathbf{C2}$. Moreover, by induction on the 
		structure of $\varphi$,
		it is straightforward to
		prove that 
		``$(M,v) \models \varphi$ iff $(M',v) \models \varphi$''
		for every $v \in W$.
		The crucial point of the proof
		is that 
		$\sim_{\allins}' = \sim_{\allins}$ and $\sim_{\univ}' = \sim_{\univ}$.
		Thus, $(M',w_0) \models \varphi$
		since $(M,w_0) \models \varphi$.\qed
	\end{proof}

	The second result is that satisfiability
	for formulas in 
	$\mathcal{L} $
	relative to the class 
	$\mathbf{QMDM}$
	is equivalent to
	satisfiability
	relative to the class
	finite-$\mathbf{QMDM}$.
	\begin{theorem}\label{theo:equiv3}
		Let $\varphi \in \mathcal{L}  $.
		Then, $\varphi$
		is satisfiable relative to the
		class 
		$\mathbf{QMDM}$
		if and only if it is satisfiable
		relative
		to the class  finite-$\mathbf{QMDM}$.
	\end{theorem}
	\begin{proof}
		The right-to-left direction is clear.
		We are going to prove the left-to-right direction
		by a filtration argument.

		Let 
		$M  =  (W, \sim_{\allins}, \sim_{\univ}, V)$ be a quasi-MDM
		and $w_0 \in W$
		such that $(M,w_0) \models \phi $.
		It is routine to verify that 
		$(  \sim_{\allins} \cup \sim_{\univ})^*= 
		\sim_{\allins} \circ \sim_{\univ}=
		\sim_{\univ} \circ  \sim_{\allins} $.
		Thus, we can define 
		$M'  =  (W', \sim_{\allins}', \sim_{\univ}', V')$
		to be the submodel
		of $M$
		generated from $w_0$
		through the relation $\sim_{\allins} \circ \sim_{\univ}$. 
		$M'$
		is a quasi-MDM
		and $(M',w_0) \models \phi $.

		Let $\mathit{sf}(\phi)$ be the set of all subformulas of $\phi$
		and  let
		$\mathit{sf}^+(\phi)=\mathit{sf}(\phi) \cup   \dec$.
		Moreover, for every $v \in W'$, let $\Theta( v )=\big\{ \psi \in
		\mathit{sf}^+(\phi) : (M',v) \models \psi 
		\big\}$.
		For every $v,u \in W'$,
		we define 
		\begin{align*}
			v  \simeq u \text{ iff }& \Theta( v )=\Theta( u ). 
		\end{align*}    
		Moreover,
		we define 
		$[v ] = \{ u \in W' :
		v  \simeq u
		\}$. 
		
		We construct a new model 
		$M''  =  (W'', \sim_{\allins}'', \sim_{\univ}'', V'')$
		where:
		\begin{itemize}
			\item[-] $W''= \{[v ] : 
			v \in W'  \}$;
			
			\item[-]
			$[v ]  
			\sim_{\allins}'' [u ] $
			iff 
			\begin{align*}
				\forall \allins
				\psi \in 
				\mathit{sf} (\phi),
				\big( (M',v) \models 
				\allins
				\psi \text{ iff } 
				(M',u) \models 
				\allins
				\psi \big);
			\end{align*}

			\item[-]
			$[v ] 
			\sim_{\univ}'' [u ] $
			iff 
			\begin{align*}
				& \forall \univ 
				\psi \in 
				\mathit{sf} (\varphi ) ,
				\big( (M',v) \models 
				\univ 
				\psi \text{ iff } 
				(M',u) \models 
				\univ 
				\psi \big) \text{ and }\\
				& \forall p 
				\in 
				\mathit{sf}(\phi) \cap
				\atm_0 ,
				\big( (M',v) \models 
				p \text{ iff } 
				(M',u) \models 
				p \big) ;
			\end{align*}

			\item[-] 
			$V'' \big(
			[v ]  \big) =
			V_{ \mathit{sf} (\phi) \cap \atm_0 }' (v)
			\cup V_\dec'(v)$.  
			
		\end{itemize}

		$M''$ is indeed a filtration, for it satisfies that if $v \sim_{\blacksquare} u$, then $[v] \sim_{\blacksquare} [u]$; and if $\blacksquare \psi \in sf(\phi)$ and $(M', v) \models \blacksquare \psi$, then $(M', u) \models \psi$, for $\blacksquare \in \{\allins, \univ\}$.
		Additionally, the valuation function is defined in the standard way.
		
		To check that $M''$ is a finite quasi-MDM, we go through all constraints.
		For $\mathbf{C1}$ a crucial fact is that $M'$ generated from $w_0$ through $\sim_{\allins} \circ \sim_{\univ}$, viz. $\forall v, u \in W', v \sim_{\allins} \circ \sim_{\univ} u$ and $v \sim_{\univ} \circ \sim_{\allins} u$.

		To see that fact, by construction of $M'$ we have $w_0 \sim'_{\allins} \circ \sim'_{\univ} u$ and $w_0 \sim'_{\allins} \circ \sim'_{\univ} v$. This means $w_0 \sim'_{\allins} v_1 \sim'_{\univ} v$ and $w_0 \sim'_{\allins} u_1 \sim'_{\univ} u$ for some $u_1, v_1 \in W' \subseteq W$.
		Then we have $v_1 \sim'_{\allins} \circ \sim'_{\univ} u$ by the Euclidean of $\sim'_{\allins}$.
		Then, since $v_1, u \in W$ and by $\mathbf{C1}$ of $M$, we have $v_1 \sim_{\univ} v_2 \sim_{\allins} u$ for some $v_2 \in W$. Since $w_0 \sim_{\allins} v_1 \sim_{\univ} v_2$, we are sure that $v_2 \in W'$, which gives us $v_1 \sim'_{\univ} v_2 \sim'_{\allins} u$. So now we have $v \sim'_{\univ} v_1 \sim'_{\univ} v_2 \sim'_{\allins} u$, and by Euclidean of $\sim'_{\univ}$, we have $v \sim'_{\univ} \circ \sim'_{\allins} u$. The case of $u \sim'_{\allins} \circ \sim'_{\univ} v$ is proven in the same way.
		
		$\mathbf{C3}$ holds because of the definition of $\sim_{\univ}''$.
		$\mathbf{C4}, \mathbf{C5}$ hold, since $V''$ not only considers $\mathit{sf}(\phi) \cap \atm_0$ but also $\dec$.
		
		It is routine to verify that 
		$M''  =  (W'', \sim_{\allins}'', \sim_{\univ}'', V'')$ is a filtration
		of $M'$ and is a finite quasi-MDM. 
		Therefore, $(M'',[w_0]) \models \varphi$.\qed    
	\end{proof}

	The following 
	theorem
	is provable
	by standard canonical model argument. 
	Note that like Theorems 
	\ref{theo:equiv1} and     \ref{theo:equiv3}, it
	does not rely on 
	$\atm_0$
	being infinite or finite.
	\begin{theorem}\label{theo:complInf}
		The logic $\wbclu$ is sound and complete relative
		to the class 
		$\mathbf{QMDM}$.
	\end{theorem}
	
	The fact that the logic 
	$\wbclu$
	is sound and complete relative to the class 
	$\mathbf{MCM}$
	is a direct corollary
	of Theorems 
	\ref{theo:equiv1},
	\ref{theo:equiv2},
	\ref{theo:equiv3} and 
	\ref{theo:complInf}. 
	
	\begin{corollary}\label{coro:infincompl}
		Let $\atm_0 $
		be infinite.
		Then,
		the logic $\wbclu$ is sound and complete relative
		to the class 
		$\mathbf{MCM}$.
	\end{corollary}

	\subsection{Complexity Results}

	We now move to  complexity of satisfiability checking.
	As for the axiomatics, we distinguish
	the finite-variable from
	the infinite-variable case.
	When 
	$\atm_0 $ is finite,
	the problem
	of verifying whether a formula
	is satisfiable is polynomial.
	The latter problem 
	mirrors 
	the 
	satisfiability
	checking problem for the finite-variable
	modal logic S5 
	which is also  known to be polynomial \cite{DBLP:journals/ai/Halpern95}.

	\begin{theorem}\label{theo:compl1}
		Let $\atm_0 $
		be finite. 
		Then,
		checking satisfiability of $\mathcal{L} $-formulas
		relative to the class
		$\mathbf{MCM}$  can be done in polynomial time.
	\end{theorem}
	\begin{proof}
		Suppose
		$|\atm_0| $ is finite.
		Then, the class
		$\mathbf{MCM}$
		is bounded by some integer $k $. 
		So, in order to determine
		whether a formula $\varphi$
		is satisfiable
		for the class 
		$\mathbf{MCM}$,
		it is sufficient to verify whether
		$\varphi $
		is satisfied in one of these MCMs. This verification
		takes a polynomial time in the size of
		$\varphi $ since it is a repeated model checking 
		and model checking in the product modal logic S5$^2$
		is polynomial.\qed
	\end{proof}
	We know that 
	when moving from the finite-variable
	to the infinite-variable
	case
	complexity of satisfiability checking
	is in NEXPTIME.
	\begin{theorem}\label{theo:compl2}
		Let $\atm_0 $
		be infinite.
		Then, checking satisfiability of $\mathcal{L} $-formulas
		relative to the class
		$\mathbf{MCM}$  is in NEXPTIME.
	\end{theorem}
	\begin{proof}
		We know that satisfiability
		checking
		for
		the product modal logic S5$^2$
		with two S5 modalities $\Box_1$
		and $\Box_2$
		is NEXPTIME-complete \cite{GabbayWolther}. 
		We have a polynomial reduction of satisfiability checking for
		$\mathcal{L} $-formulas
		relative to the class
		$\mathbf{MCM}$ 
		to the latter problem. 
		In particular, 
		given a formula $\varphi \in \mathcal{L} $,
		we can translate it into
		a formula 
		$\mathit{tr}(\varphi)$
		of S5$^2$
		where the translation 
		$\mathit{tr}$
		is defined as follows:
		(i) $\mathit{tr}(q)=q$
		for $q \in \mathit{Atm}$,
		(ii) $\mathit{tr}(\neg \varphi)=\neg \mathit{tr}(\varphi)$,
		(iii) $\mathit{tr}(\varphi_1 \wedge \varphi_2)= \mathit{tr}(\varphi_1) \wedge \mathit{tr}(\varphi_2)$,
		(iv)
		$\mathit{tr}(\allins \varphi)= \Box_1\mathit{tr}(\varphi) $,
		(v)
		$\mathit{tr}(\univ \varphi)= \Box_2\mathit{tr}(\varphi) $. 
		We have that $\varphi$
		is satisfiable for
		the class
		$\mathbf{MCM}$
		if and only 
		$\bigwedge_{\chi \in \Delta} 
		\Box_1\Box_2
		\chi
		\wedge \mathit{tr}( \varphi) $
		is 
		a satisfiable 
		formula
		of the product modal logic S5$^2$,
		where $\Delta$ is the following finite theory 
		corresponding to the Axioms 
		\ref{ax:Ind1}, \ref{ax:Ind2},
		\ref{ax:AtMost} and
		\ref{ax:AtLeast}
		of the logic $\wbclu$:
		\begin{align*}
			\Delta=&\{\bigvee_{x \in \val} \takevalue{x}\} \cup
			\{\takevalue{x} \to \neg \takevalue{y} :
			x\neq y \} \cup
			\{ p \rightarrow {\univ}  p :
			p \in \mathit{Atm}_0(\varphi)  \} \cup\\
			&     \{ \neg p \rightarrow {\univ}  \neg p :
			p \in  \mathit{Atm}_0(\varphi)  \},
		\end{align*}
		and $ \mathit{Atm}_0(\varphi)$
		is the set of atoms in $ \mathit{Atm}_0(\varphi)$
		which occur in $\varphi$. \qed
	\end{proof}
	In
	\cite{DBLP:journals/sLogica/BezhanishviliH04}
	(see also \cite{DBLP:journals/sLogica/BezhanishviliM03})
	it is proved that
	all proper normal
	extensions of the product modal 
	logic
	S5$^2$ are in NP. 
	In future work, we plan
	to verify whether these results
	are applicable to our setting
	in order 
	to improve our complexity
	upper bound. 
	The problem is that Axioms
	\ref{ax:Ind1},
	\ref{ax:Ind2},
	\ref{ax:AtMost} and
	\ref{ax:AtLeast}
	are not axiom schemata
	in the proper sense. 

	\section{Application}
	
	As mentioned, the ${\univ}$ operator is interpreted as partial knowledge 
	about the classifier properties.\footnote{In the real world, partial knowledge may come from the data set as
		well as from the training process.
		For example, through learning, 
		we may acquire knowledge that certain input features behave monotonically \cite{you2017deep}. } 
	In this section, we are going to exemplify how to
	use it for representing   abductive explanations of
	a black box classifier. 
	
	
	\subsection{An Example of Classification Task}\label{sec:paperexa}

	Consider   a selection function 
	which specifies whether a paper submitted to a conference is acceptable  for presentation 
	(1) or not (0) depending on its feature profile
	composed of four input features: significance ($\mathrm{si}$), originality ($\mathrm{or}$), clarity of the presentation ($\mathrm{cl}$)
	and  fulfillment of the anonymity requirement ($\mathrm{an}$). 
	For the sake of simplicity,
	we assume each feature in a paper profile is binary: $\mathrm{si}$ means
	the paper is significant while $\neg \mathrm{si}$
	means the paper is not significant,
	$\mathrm{or}$ means
	the paper is original while $\neg \mathrm{or}$
	means the paper is not original, and so on.
	We say that  a first paper profile dominates a second paper profile, if all conditions
	satisfied by the second profile are satisfied by the first profile,
	and there exists a condition satisfied by the first profile which
	is not satisfied by the second profile. 
	For example if the first profile is $\mathrm{si} \wedge
	\neg \mathrm{or} \wedge \mathrm{cl} \wedge \mathrm{an}$
	and the second profile is
	$\mathrm{si} \wedge
	\neg \mathrm{or} \wedge \neg \mathrm{cl} \wedge \mathrm{an}$, then the first dominates the second.
	
	The selection function is implemented in a classifier
	system 
	that has to automatically split papers
	into two sets, the set of acceptable papers and the set of non-acceptable ones.  
	We assume a certain agent (e.g., the author
	of a paper submitted to the conference)
	has only partial knowledge of the classifier system. 
	In particular,
	she   only knows that the classifier
	complies with the following three constraints:
	(1) submissions that satisfy  the four conditions should be automatically accepted,
	(2) if a first paper profile dominates a second paper profile and the second paper profile
	is acceptable, then the first paper profile should also be acceptable,
	and (3) submissions that violate the anonymity requirement should be automatically rejected.
	In this case, the classifier is a black box
	for the agent.

	\begin{example}\label{ex: paper}
		The multi-classifier model (MCM) representing
		the previous situation 
		is the  tuple 
		$\mcm$
		such that
		$S= 2^{ \{
			\mathrm{si} 
			,\mathrm{or} , \mathrm{cl} , \mathrm{an}
			\}  }$
		and 
		\begin{align*}
			\forall f \in F_S,
			f \in \Phi \text{ iff } 
			(i) & \
			\forall s \in S,
			\text{ if } 
			\{
			\mathrm{si} 
			,\mathrm{or} , \mathrm{cl} , \mathrm{an}
			\}\subseteq s 
			\text{ then } f(s)=1 ,\\
			(ii) & \
			\forall s,s' \in S, \text{ if } 
			s \subset s' \text{ and }
			f(s)=1 
			\text{ then } 
			f(s')=1. \\
			(iii) & \
			\forall s \in S,  
			\text{ if } 
			\mathrm{an} \not  \in s
			\text{ then } f(s)=0.
		\end{align*}
		The agent does not know which
		function 
		in $\Phi $ corresponds to the actual
		classifier, i.e., 
		they are epistemically indistinguishable for her.
	\end{example}

	\subsection{Explanations}
	Given space constraints, we exemplify explanations for white and black box classifiers by showing the dichotomy global vs. local explanation and the notion of \emph{abductive explanation} based on \emph{prime implicant}.
	Some notations and abbreviations  are needed to formally represent them.
	Let $\lambda$ denote a conjunction of literals, where a literal is an atom $p$ or its negation $\neg p$.
	We write $\lambda \subseteq \lambda'$, call $\lambda$ a part (subset) of $\lambda'$, if all literals in $\lambda$ also occur in $\lambda'$; and $\lambda \subset \lambda'$ if $\lambda \subseteq \lambda'$ but not $\lambda' \subseteq \lambda$.
	In the glossary of Boolean classifiers, $s$ is called an \emph{instance}, $\lambda$ is called a \emph{term} or \emph{property} (of the instance).
	The set of terms
	is noted $ \mathit{Term}$. 
	Moreover, let $\atm(\phi)$ denote the atoms occurring in $\phi$.
	Finally, notice that the abbreviations $[X]\phi$ and $\langle X \rangle \phi$ introduced in Section 2 will be used.
	
	
	Let us start with \emph{prime implicant}, a key concept in
	the theory of Boolean functions since \cite{quine1955way}.
	It can be presented in the language
	$\mathcal{L} (\atm)$
	as follows:
	\begin{align*}
		\pimp(\lambda, x) =_{\mathit{def} } & \allins \Big( \lambda \to \big(\takevalue{x} \wedge 
		\bigwedge_{p \in \atm(\lambda)} \langle \atm(\lambda)\setminus\{p\}\rangle \neg \takevalue{x}  \big) \Big).
	\end{align*}
	The abbreviation $    \pimp(\lambda, x)$ has to be
	read  ``$\lambda$ is a prime implicant for the classification $x$''.
	Roughly speaking, the latter means that 
	(i) 
	$\lambda$ necessarily leads to
	the classification $x$ (why $\lambda$ is an \emph{implicant}), and 
	(ii) for any
	of 
	its proper subsets $\lambda'$, possibly there is a state where $\lambda'$ holds but 
	the classification 
	is different from $x$ (why $\lambda$ is \emph{prime}).

		
	
	
	Prime implicant counts as a ``global'' explanation, in the sense that it is a property of the classifier and holds at \textit{all} its input instances.
	Partially,  as a response  to the local approach in model-agnostic methods, researchers from logic-based approaches in XAI focus on the ``localized'' prime implicant namely \textit{abductive explanation} (AXp).\footnote{ It has many names in literature: PI explanation \cite{shih2018formal}, sufficient reason \cite{DBLP:conf/ecai/DarwicheH20}. We adopt the one from \cite{ignatiev2019abduction} for its nice correspondence to contrastive explanation in \cite{ignatiev2020contrastive}. }
	An abductive explanation
	is not only a prime implicant, but also  a \emph{property of the actual instance}.
	The notion of abductive explanation
	is expressed in
	$\mathcal{L} $ as follows: 
	\begin{align*}
		\axp(\lambda, x)
		=_{\mathit{def} } 
		\lambda \wedge \pimp(\lambda, x).
	\end{align*}
	$     \axp(\lambda, x)$
	just means that 
	$\lambda$ 
	is an abductive explanation of the actual
	classification  $x$. 
	Let us instantiate the notions
	of prime implicant and abductive explanation
	in the paper example we introduced in Section \ref{sec:paperexa}.
	
	\begin{example}\label{ex: paper 2}
		Take the MCM \mcm\ in Example \ref{ex: paper}, and let $s_1 = \{\mathrm{si}, \mathrm{or}, \mathrm{an}\} \in S$.
		Consider the function $f_1$ 
		s.t.
			$\forall s \in S:  f_1(s)= 1 \text{ iff }
			\mathrm{an}
			\in  s 
			\text{ and }
			\{\mathrm{or}, \mathrm{cl}\}
			\cap  s  \neq \emptyset  
			$.
		The function $f_1$
		is syntactically expressed
		by the formula   $\allins \big(\takevalue{1} \leftrightarrow ((\mathrm{or} \wedge \mathrm{an}) \vee (\mathrm{cl} \wedge \mathrm{an})) \big)$. Clearly $f_1 \in \Phi$ for it satisfies the three constraints. Hence, we have:
		\begin{align*}
			(\Gamma, s_1, f_1) \models & \axp(\mathrm{or}\wedge\mathrm{an}, 1) \wedge \pimp(\mathrm{or}\wedge\mathrm{an}, 1) \wedge \pimp(\mathrm{cl}\wedge\mathrm{an}, 1).
		\end{align*}
		Meanwhile $(\Gamma, s_1, f_1) \not \models  \axp(\mathrm{cl}\wedge\mathrm{an}, 1)$, because $(\Gamma, s_1, f_1) \not \models  \mathrm{cl}\wedge\mathrm{an}$.
		But consider $s_2 = \{\mathrm{si}, \mathrm{cl}, \mathrm{an}\} \in S$.
		We have $(\Gamma, s_2, f_1) \models \axp(\mathrm{cl}\wedge\mathrm{an}, 1)$.
	\end{example}
	

	Now we investigate what happens when facing a black box model \mcm. 
	The agent has uncertainty about the actual classifier's properties.
	Therefore, it is interesting to draw
	the distinction between objective
	and subjective (or epistemic) explanation. 
	Objective explanation coincides with the notion
	of explanation
	in the context of white box classifiers
	defined above. 
	Subjective explanation refers 
	to the agent's 
	interpretation
	of the classifier 
	and her  explanation 
	of the classifier's decision
	in the light of her partial knowledge.

	We say the term
	$\lambda$
	is a \emph{subjective}
	prime implicant 
	for $x$, noted $\mathtt{SubPImp}(\lambda,x)$, 
	if the agent knows
	that 
	$\lambda$
	is a prime implicant for $x$, that is:
	\begin{align*}
		\mathtt{SubPImp}(\lambda,x)  =_{\mathit{def}} 
		{\univ} \pimp(\lambda, x).
	\end{align*}
	Similarly, we say $\lambda$
	is
	a \emph{subjective}
	abductive explanation of the actual
	classification  $x$, noted $\mathtt{SubAXp}(\lambda,x)$,
	if the agent knows that 
	$\lambda$ is 
	an 
	abductive explanation of the actual
	classification  $x$,
	that is:
	\begin{align*}
		\mathtt{SubAXp}(\lambda,x)  =_{\mathit{def}} 
		{\univ} \axp(\lambda, x).
	\end{align*}

	It is worth noting that 
	in the case of a white box classifier, 
	if the set of input
	instances $S$
	is finite, 
	we can always find 
	an abductive explanation
	of the actual classification. That is,
	for every $ \mcm \in \mathbf{MCM} $, $s \in S$ and $f \in \Phi $:
	\begin{align*}
		\text{ if } 
		S \text{ is finite then }
		\exists \lambda \in \mathit{Term}
		\text{ such that }
		(\Gamma, s, f) \models \axp\big(\lambda, f(s) \big).
	\end{align*}
	Nonetheless, this result  cannot
	be generalized
	to the black box case.
	Indeed, as
	the following example shows,
	there is no guarantee for
	the existence of a subjective explanation of the actual
	classification. 
	The problem is that the minimality condition
	can collapse when moving from
	objective to subjective explanation,
	since the agent can have more than one classifier
	in her epistemic state. 
	\begin{example}
		Let $\mcm$, $f_1$ and $s_1$ be the same as
		in Example \ref{ex: paper 2}. 
		There is no $\lambda$ such that  $(\Gamma, s_1, f_1) \models {\univ} \axp(\lambda, 1)$. 
		To see this, consider $f_2$ 
		s.t. 
			$      \forall s \in S:  f_2(s)= 1 \text{ iff }
			\{\mathrm{si}, \mathrm{an}\}
			\subseteq   s   .
			$
		The function $f_2$
		is syntactically expressed
		by the formula   $\allins (\takevalue{1} \leftrightarrow (\mathrm{si} \wedge \mathrm{an}) )$.
		Clearly $f_2 \in \Phi$ for it satisfies the three constraints.
		We have $(\Gamma, s_1, f_2) \models \axp(\mathrm{si} \wedge \mathrm{an}, 1)$.
		But there is no term which  minimally explains both
		$f_1(s_1)$
		and $f_2(s_1)$.
		Indeed, 
		$\mathrm{or} \wedge \mathrm{an}$ is not enough for
		explaining 
		$f_2(s_1)$, $\mathrm{si} \wedge \mathrm{an}$
		is not enough for explaining
		$f_1(s_1)$, and $\mathrm{si} \wedge \mathrm{or} \wedge \mathrm{an}$ fails the minimality condition for both.
		Therefore, we have
		\begin{align*}
			(\Gamma, s_1, f_1) \models &  \axp(\mathrm{or}\wedge\mathrm{an}, 1) \wedge 
			\bigwedge_{ \lambda \in \mathit{Term}(
				\{\mathrm{si}, \mathrm{or},\mathrm{cl}, \mathrm{an}\}
				)  }
			\neg \mathtt{SubAXp}(\lambda, 1).
		\end{align*}
		However, this does not mean that the agent knows
		nothing about the classifier.
		For instance,
		she knows that violating the anonymity 
		requirement is a prime implicant
		for rejection, that is,
		$(\Gamma, s_1, f_1) \models  \mathtt{SubAXp}(\neg \mathrm{an}, 0)$.
	\end{example}
	
		
	
	To sum up, the four notions of explanation we introduced 
	can be organized in Table
	\ref{tab:notions}
	along the two dimensions 
	objective vs
	subjective
	and local vs global. 
	\begin{table}[h]
		\centering 
		\begin{tabular}{|l|c|c|}
			\hline
			& Local & Global \\
			\hline
			Objective & $\axp(\lambda, x)$ & $\pimp(\lambda, x)$ \\
			\hline
			Subjective & $\mathtt{SubAXp}(\lambda, x)$ & $\mathtt{SubPImp}(\lambda, x)$ \\
			\hline
		\end{tabular}
		\caption{Notions of prime implicant 
			and abductive explanation}
		\label{tab:notions}
	\end{table}
	

	\section{Dynamic Extension}\label{dynext}
	
	Before concluding,
	we are going to present a simple dynamic
	extension
	of the language $\mathcal{L} $
	by operators of the form
	$[ \varphi ]$.
	They describe 
	the consequences  of removing from
	the actual
	model
	all classifiers that 
	do not \emph{globally} satisfy the constraint $\varphi$.
	More generally, they allow
	us to model
	the process 
	of 
	gaining new knowledge about the
	classifier's properties. 
	The extended modal language  $\mathcal{L}^{\mathit{dyn}}  $
	is defined by the following grammar:
	\begin{center}\begin{tabular}{lcl}
			$\varphi$  & $::=$ & $ p \mid
			\takevalue{x} \mid 
			\neg\varphi \mid \varphi_1\wedge\varphi_2 \mid \allins\varphi \mid {\univ} \phi
			\mid [ \varphi  ] \psi ,$
	\end{tabular}\end{center}
	where $p$ ranges over $ \atm_0 $
	and 
	$x $
	ranges over $\val $.
	
	The new formula
	$ [ \varphi  ] \psi$
	has to be read ``$\psi$
	holds
	after having discarded all classifiers
	that do not globally satisfy
	the property  $\varphi$''. 
	Notice the similar but different notations $[X]$ and $[\phi]$. For example, $[\{p\}], [\{p, q\}]$ are abbreviations with ceteris paribus meaning, while $[p], [p \wedge \neg q]$ are dynamic operators.
	
	The interpretation
	of the operators 
	$[ \varphi  ] $
	relative to a pointed MCM
	$ ( \Gamma ,  s,f  ) $
	with 
	$\mcm$, 
	$s \in S$
	and $f \in \Phi$
	goes as follows:
	\begin{eqnarray*}
		( \Gamma ,  s,f  ) \models 
		[ \varphi  ] \psi & \iff &
		\text{if }  
		( \Gamma ,  s,f  ) \models \allins \varphi 
		\text{ then }
		( \Gamma^\phi  ,  s,f  ) \models \psi,
	\end{eqnarray*}
	where 
	$\Gamma^\phi= (S^\phi, \Phi^\phi) $
	is the MCM
	such that: 
	\begin{align*}
		& S^\phi = S,\\
		& \Phi^\phi= \{f' \in \Phi :
		\forall s' \in S, 
		( \Gamma  ,  s' ,f'  )\models 
		\phi
		\}. 
	\end{align*}
	The previous update
	semantics for the 
	operator
	$  [ \varphi  ]$
	is reminiscent of the semantics
	of public announcement logic (PAL) \cite{Plaza2007,kooietalDEL2007}.
	However, there is an important difference.
	While  PAL has a 
	one-dimensional 
	state
	elimination semantics,
	our update semantics operates
	on a single dimension of
	the product in an MCM.
	In particular, it only removes
	classifiers
	that do not globally
	satisfy the constraint $\varphi$,
	without modifying the set $S$ of input
	instances. 
	
	The logics
	$\dbclu$ and 
	$\dwbclu$ (Dynamic $\bclu$ and $\dwbclu$)
	extend the logic
	$\bclu$
	and 
	$\wbclu$
	by the dynamic
	operators $  [ \varphi  ] $.
	They are defined as follows.

	\begin{definition}[Logics $\dbclu$ and
		$\dwbclu$]\label{axiomatics3}
		We define  $\dbclu$
		(resp. $\dwbclu$)
		to be the extension of  $\bclu$
		(resp. $\wbclu$)
		of Definition \ref{axiomatics}
		(resp. Definition \ref{axiomatics2})
		generated by the following reduction axioms for the dynamic operators 
		$  [ \varphi  ] $:
		\begin{align*}
			[ \varphi  ]p  \leftrightarrow&  ( \allins \varphi
			\rightarrow p ) \\
			[ \varphi  ]\takevalue{x} \leftrightarrow&   \big(  \allins\varphi
			\rightarrow \takevalue{x} \big) \\
			[ \varphi  ]\neg \psi   \leftrightarrow&   ( \allins\varphi
			\rightarrow  \neg  [ \varphi  ] \psi)   \\
			[ \varphi  ] (\psi_1 \wedge \psi_2) \leftrightarrow& 
			\big( [ \varphi  ] \psi_1 \wedge
			[ \varphi  ] \psi_2 \big)\\
			[ \varphi  ] \allins
			\psi\leftrightarrow &
			( \allins \varphi
			\rightarrow   \allins[ \varphi  ] 
			\psi )\\
			[ \varphi  ] \univ 
			\psi\leftrightarrow &
			( \allins \varphi
			\rightarrow   \univ [ \varphi  ] 
			\psi )
		\end{align*}
		and the following rule of inference:
		\begin{align}
			&  \frac{\varphi_1 \leftrightarrow \varphi_2}{
				\psi \leftrightarrow \psi[\varphi_1/\varphi_2]
			} \tagLabel{RE}{ax:rulere} 
		\end{align}
	\end{definition}

	It is a routine
	exercise
	to verify that 
	the equivalences
	in Definition \ref{axiomatics3}
	are valid for the class
	$\mathbf{MCM}$
	and that the rule of replacement of equivalents 
	(\ref{ax:rulere})
	preserves validity.
	We show the validity of the sixth
	equivalence as an example:
	\begin{align*}
		( \Gamma ,  s,f  ) \models 
		[ \varphi  ] \univ 
		\psi    \Longleftrightarrow &
		\text{ if }  
		( \Gamma ,  s,f  ) \models \allins \varphi 
		\text{ then}
		( \Gamma^\phi  ,  s,f  ) \models 
		\univ 
		\psi;\\
		\Longleftrightarrow &
		\text{ if }  
		( \Gamma ,  s,f  ) \models \allins \varphi 
		\text{ then}
		\forall f' \in \Phi^\phi, 
		( \Gamma^\phi  ,  s,f'  ) \models  \psi;\\
		\Longleftrightarrow &
		\text{ if }  
		( \Gamma ,  s,f  ) \models \allins \varphi 
		\text{ then }   \forall f' \in \Phi,\\
		&   \big( \text{if }
		\forall s' \in S,
		( \Gamma  ,  s',f'  ) \models  \psi
		\text{ then } ( \Gamma^\phi  ,  s,f'  ) \models  \psi \big);\\
		\Longleftrightarrow &
		\text{ if }  
		( \Gamma ,  s,f  ) \models \allins \varphi 
		\text{ then }   \forall f' \in \Phi,\\
		& \
		\big( \text{if }
		( \Gamma  ,  s,f'  ) \models \allins    \psi
		\text{ then } ( \Gamma^\phi  ,  s,f'  ) \models  \psi \big);\\
		\Longleftrightarrow &
		\text{ if }  
		( \Gamma ,  s,f  ) \models \allins \varphi 
		\text{ then }   \forall f' \in \Phi,   ( \Gamma  ,  s,f'  ) \models [\varphi]    \psi; \\
		\Longleftrightarrow &
		( \Gamma ,  s,f  ) \models \allins \varphi 
		\rightarrow \univ[\varphi]    \psi .
	\end{align*}

	The completeness of $\dbclu$
	and $\dwbclu$
	for this class of models
	follows from Theorem 
	\ref{theor: Completeness Finite}
	and Corollary 
	\ref{corollComp}, in view of the fact that the reduction axioms
	and the rule of replacement of proved
	equivalents
	can be used to find, for any 
	$\mathcal{L}^{\mathit{dyn}} $-formula, a provably equivalent 
	$\mathcal{L} $-formula.
	
	\begin{theorem}
		
		Let $\atm_0 $
		be finite.
		Then, 
		the logic 
		$\dbclu$
		is sound and complete relative
		to the class $\mathbf{MCM}$.
	\end{theorem}
	
	\begin{theorem}
		Let $\atm_0 $
		be infinite.
		Then, 
		the logic 
		$\dwbclu$
		is sound and complete relative
		to the class $\mathbf{MCM}$.
	\end{theorem}
	
	The following decidability result
	is a consequence of
	Theorem
	\ref{theo:compl2}
	and the fact that
	via 
	the reduction axioms
	in Definition \ref{axiomatics3}
	we
	can find
	a 
	reduction
	of satisfiability
	checking of $\mathcal{L}^{\mathit{dyn}} $-formulas
	to 
	satisfiability
	checking  of $\mathcal{L} $-formulas.
	\begin{theorem}\label{theo:compldyn}
		Checking satisfiability
		of $\mathcal{L}^{\mathit{dyn}}$-formulas 
		relative to 
		$\mathbf{MCM}$ 
		is decidable.
	\end{theorem}
	
	
	
	Let us end up with the paper example
	to illustrate to expressive power of our dynamic extension.
	\begin{example}\label{ex: paper 3}
		Let $\mcm, f_1$ and $s_1$ be the same as in Example \ref{ex: paper 2}.
		We have
		\begin{align*}
			(\Gamma, s_1, f_1) \models [(\mathrm{or} \wedge \mathrm{an}) \to \takevalue{1}] \univ \bigvee_{\lambda \subseteq (\mathrm{or} \wedge \mathrm{an})} \axp(\lambda, 1).
		\end{align*}
		This means that after having discarded all classifiers which do not take $(\mathrm{or} \wedge \mathrm{an})$ as an implicant for
		acceptance of a paper, the agent knows that there must be a part of $\mathrm{or} \wedge \mathrm{an}$ that abductively explains the acceptance of the paper $s_1$.
	\end{example}
	

	\section{Conclusion}

	We have presented a
	product modal logic  
	which supports
	reasoning about (i)
	partial knowledge
	and 
	uncertainty
	of a
	classifier's  properties
	and,
	(ii)  objective and subjective explanations 
	of a classifier's decision. 
	Moreover, we have studied 
	a dynamic extension 
	of the logic which allows us to represent 
	the event  of 
	gaining new knowledge about the classifier's properties.

	Our logic
	is intrinsically single-agent:
	it models the uncertainty
	of one agent about the actual 
	classifier's properties. In future work,
	we plan to generalize
	our framework to the multi-agent setting.
	The extension would result
	in a multi-relational product semantics
	in which every agent has her own
	epistemic indistinguishability
	relation which commutes with the 
	input instance dimension 
	(the equivalence
	relation $\sim_{\allins}$
	in Definition \ref{MDM} of MDM). 
	We also plan to enrich this semantics
	with a knowledge update mechanism
	in the spirit of Section 
	\ref{dynext}. This 
	would allow us to represent 
	exchange of information 
	between agents
	with an explanatory purpose, 
	which is named dialogical explanation by philosophers \cite{Walton2004} and interactive explanation by researchers in the XAI domain \cite{AmershiETAL,Miller2019}.

	
	\section*{Acknowledgments }
	Support from the ANR-3IA Artificial and Natural Intelligence Toulouse Institute 
	(ANITI) is gratefully acknowledged.

	\bibliographystyle{splncs04}
	\bibliography{biblio}

\end{document}